\documentstyle[prl,aps,twocolumn,epsfig]{revtex}

\begin{document}

\twocolumn[\hsize\textwidth\columnwidth\hsize\csname @twocolumnfalse\endcsname
\title{Possible evidence for a change in cosmic equation of state at
decoupling}

\author{Julio A. Gonzalo and Manuel I. Marqu\'es}

\address {Depto. de F\'isica de Materiales C-IV
Universidad At\'onoma de Madrid
Cantoblanco 28049 Madrid Spain}

\date{19-1-00}

\maketitle

\begin{abstract}
Indirect but significative evidence (from fusion research data,
and from the known $^4He$ abundance) relevant to primordial nucleosynthesis
is examined. It is shown that this evidence supports a change in cosmic
equation of state at decoupling, compatible with Einstein's cosmological
equations, from $\rho _mT^{-4}\approx const.(\rho _m=\rho _r)$ prior to
decoupling $(t<t_{af})$, to $\rho _mT^{-3}\approx const.$ after $(t>t_{af})$,
being $t_{af}$ the atom formation time. Implications at the baryon
threshold time (temperature) and beyond are discussed.\hspace{5em}\vskip1pcPACS numbers: 05.70.Ce, 040.20.Me, 98.80.-k
\end{abstract}

\vskip1pc]

\narrowtext


{\bf I. Introduction} \vskip1pc Recent studies of distant Type 1a supernovae
indicate that the universe will expand forever \cite{Physics}. According to
them the universe does not have enough matter (visible or dark) to halt the
current expansion. The new findings are consistent with an estimate of the
time elapsed since the big-bang of the order of $15\times 10^9$years.

The usual (big-bang) approach to cosmic evolution has taken for granted
during decades that a single equation of state, $\rho _{m}T^{-3}=const.$, is
capable of describing well cosmic dynamics at all times. Since at the time
of atom formation (last scattering) the physical contents of the universe
underwent a drastic change from a plasma to a gas of neutral atoms, the
validity of the same equation of state prior to and after that event might
be questionable.

In this work we examine relevant experimental evidence\cite{Cordey,Callen}
from fusion research data to check wether $\rho _{m}T^{-3}=const\ (\rho
_{m}>\rho _{r})$, where $\rho _{m}$ is the matter mass density and $\rho
_{r} $ the radiation mass density, describes well cosmic evolution between
two specific times, the time at nucleosynthesis, happened much earlier than
atom formation, and the time at atoms formation. We will show that this
experimental evidence supports $\rho _{m}T^{-4}=const.\ (\rho _{m}=\rho
_{r}) $ from early times to the atoms formation time, followed by $\rho
_{m}T^{-3}=const.$ thereafter. \vskip1pc {\bf II. Cosmological equations} %
\vskip1pc We summarize the parametric Friedmann-Lemaitre solutions $(\Lambda
=0)$ of Einstein's equations and we put them in a proper from \cite{Cereceda}
for later use.

For $k\leq 0$ (open solution): 
\begin{equation}
R=R_{+}\sinh ^2y,\ t=(R_{+}/\left| k\right| ^{1/2}c)\{\sinh y\cosh y-y\}
\end{equation}

where $R_{+}$ is a constant, defined at $(8\pi G/3)\rho
_{+}R_{+}^3=c^2\left| k\right| R_{+}$. From Eq(1),

\begin{equation}
tH\equiv t(\stackrel{.}{R}/R)=\{\sinh y\cosh y-y\}\cosh y/\sinh ^3y\geq 
\frac 23,
\end{equation}

\begin{equation}
\Omega \equiv \rho /(3H^2/8\pi G)=1/\cosh ^2y
\end{equation}

For $k\geq 0$ (closed solution):

\begin{equation}
R=R_{+}\sin ^2y,\ t=(R_{+}/k^{1/2}c)\{y-\sin y\cos y\},
\end{equation}

and

\begin{equation}
tH=\{y-\sin y\cos y\}\cos y/\sin ^3y\leq \frac 23
\end{equation}

\begin{equation}
\Omega =1/\cos ^2y
\end{equation}

For $k=0$ (flat solution), $tH=2/3$, $\Omega =1$. As it is well known,
proponents of inflationary models prefer $k=0$, most observational \cite
{Schwarzschild} cosmologist agree that $0.1\leq \Omega \leq 0.3$, and $%
\Lambda \neq 0$ is not ruled out.\vskip1pc {\bf III. Equation of state for
the transparent universe} \vskip1pc There is a growing consensus \cite
{Bennet,Freedman} that the time elapsed from the big-bang is $t_o\simeq
13.7\times 10^9years$, and the present value of Hubble's constant is $%
H_o\simeq 65Km/sMpc$, resulting in

\begin{equation}
t_oH_o\simeq 0.910\pm 0.09>\frac 23
\end{equation}

( in agreement with recent studies of Type 1a supernovae).

This implies an open universe . Through Eq. (2), $y_o\simeq 1.92$ and

\begin{equation}
\Omega _o\simeq 0.0824\pm 0.0082<1.
\end{equation}

Due to the fact that at present $\rho _{mo}$(matter mass density)$>>\rho
_{ro}$(radiation mass density), we get

\begin{equation}
\rho _{mo}\simeq \Omega _o\rho _{co}\simeq \Omega _o(3H_o^2/8\pi G)\simeq
6.54\times 10^{-31}g/cm^3.
\end{equation}

This allows for a relatively moderate amount of dark matter, and even for a
certain amount of matter associated to a non-zero cosmological constant, not
enough (by an order of magnitude) to render the universe flat $(k=0)$.

Due to COBE\ data \cite{Mather} the present radiation mass density is known
with exceptional precision $(T_o=2.726\pm 0.04K)$. Then

\begin{equation}
\rho _{ro}=\frac{\sigma T_o^4}{c^2}=\frac{(7.63\times 10^{-15})(2.726)^4}{%
(3\times 10^{10})^2}=4.68\times 10^{-34}g/cm^3.
\end{equation}

In the present (transparent) universe the baryon to photon ratio, $n_b/n_r$,
remains constant in the absence of any substantial interaction between
matter and radiation. Therefore the equation of state is determined by 
\[
\frac{n_b}{n_r}=\frac{\rho _m/m_b}{(\sigma T^4)/(2.8k_BT)}=const., 
\]

resulting in

\begin{equation}
\rho _mT^{-3}=\rho _{mo}T_o^{-3}=3.23\times 10^{-32}g/cm^3K^3
\end{equation}

This equation of state should be valid at least since the time of atom
formation to present. Because $\rho _m=\rho _{mo}(R_o/R)^3$, Eq. (11)
implies $RT\approx const.$ since the time of atom formation, and, for $t<t_o$%
, the ratio $(\rho _r/\rho _m)$ was about unity at a certain time, $t_{eq}$,
at which $\rho _r=\rho _m$(equality), in the relatively distant past.\vskip%
1pc {\bf IV. Coincidence between equality and atom formation times} \vskip%
1pc At $t=t_{eq}$(equality), according to Eq.(11), we have

\begin{equation}
\rho _{meq}T_{eq}^{-3}=%
{\sigma T_{eq}^4 \overwithdelims() c^2}%
T_{eq}^{-3}=%
{\sigma  \overwithdelims() c^2}%
T_{eq},\ i.e\ T_{eq}=3808K,
\end{equation}

which implies also \cite{Cereceda} that $\rho _{meq}=1.78\times
10^{-21}g/cm^3\ $at

\begin{equation}
\ t_{eq}=1.47\times 10^{13}\sec =4.66\times 10^5yrs.
\end{equation}

On the other hand at $t=t_{af}$(atom formation), Saha's equation for $x=1/2$
(i.e. for 50\% ionized atoms), right at $T=T_{af},$ leads to \cite{Cereceda}

\begin{equation}
3.716\times 10^{-15}=%
{B_{eff} \overwithdelims() k_BT_{af}}%
^{5/2}\exp \left[ -%
{B_{eff} \overwithdelims() k_BT_{af}}%
\right] ,
\end{equation}

where $B_{eff}=14.27eV$ and $\Omega _{po}/\Omega _o=0.0355/0.0824$ may be
used, resulting in an atom formation temperature

\begin{equation}
T_{af}=3880K,
\end{equation}

in almost perfect coincidence with $T_{eq}=3808K$, given above.

This coincidence is remarkable and might suggest that $(\rho _r/\rho _m)=1$
(equality) was constant all the way from well prior to atom formation and
began to decrease towards its present value $(\rho _{ro}/\rho _{mo})\approx
7.15\times 10^{-4}$ just when the universe became transparent (we have used
the values for $\rho _{mo}$ and $\rho _{ro}$ given by Eqs. (9) and (10)
respectively). This could mean that, preceding the present matter dominated
epoch, the universe might have gone through an epoch of matter/radiation 
{\bf equality}, rather than radiation {\bf domination}, as currently assumed.

We will come back to this point after considering the evidence provided by
fusion research data and $^{4}He$ abundance.\vskip1pc {\bf V. Fusion
research data} \vskip1pc Fusion at the sun%
\'{}%
s center (complete) and primordial cosmic fusion (incomplete) admitedly take
place under different physical conditions, in particular density. However
the threshold fusion temperature is probably well represented by current
fusion research data. In what follows we will make first a rough estimate of 
$\rho _{mns}T_{ns}^{-3}$ at cosmic nucleosynthesis to be compared with $\rho
_{af}T_{af}^{-3}$ at atom formation, and we will see that the data used
suggest an equation $\rho T^{-n}\sim constant$ with $n$ different from $n=3.$
Later we will make a more refined argument using the well known $^{4}He$
abundance and using an equivalent alternative equation of state involving $t$
(time) and $T$ (temperature)$.$

The temperature and plasma density at the center of the sun \cite
{Contemporary}, where $^{4}He$ nucleosynthesis is taking place continuously,
may be expected to be within a factor of the order of a few orders of
magnitude of the cosmic nucleosynthesis temperature and density. They are
given by

\begin{equation}
T_{\odot }(ctr)\simeq 1.6\times 10^{7}K,\quad \rho _{\odot }(ctr)\simeq
150g/cm^{3}.
\end{equation}

But these data are probably not directly applicable to our problem, because
in the sun complete nucleosynthesis is produced.

Fusion research data \cite{Cordey,Callen}, on the other hand, indicate that $%
^{4}He$ synthesis is ignited (or stops) at an ion temperature given by

\begin{equation}
T_{io}=30keV=3.47\times 10^8K.
\end{equation}

What are the implications of these data for primordial nucleosynthesis?

Taking $\rho _{mns}\simeq 0.5g/cm^{3}$, $T_{ns}\simeq T_{io}$ and
introducing them in Eq. (11) one gets

\begin{equation}
\rho _{mns}T_{ns}^{-3}=1.04\times 10^{-27}g/cm^3K^3,
\end{equation}

much larger than

\begin{equation}
\rho _{mo}T_o^{-3}=\rho _{maf}T_{af}^{-3}=3.23\times 10^{-32}g/cm^3K^3,
\end{equation}

which should be expected in the absence of change in the equation of state.

On the other hand if one lets $\rho _mT^{-n}=const.,$ with $n$ free to match 
$\rho _{mns}T_{ns}^{-n}$ and $\rho _{af}T_{af}^{-n}$, one gets

\begin{equation}
n=\frac{\ln (\rho _{mns}/\rho _{af})}{\ln (T_{ns}/T_{af})}=4.12
\end{equation}

which is compatible with n=4 and therefore, with $\rho _{m}=\rho _{r}=\sigma
T^{4}/c^{2}$ prior to decoupling (atom formation). This implies

\begin{equation}
{\rho _mT^{-4}=\sigma /c^2=8.47\times 10^{-36}g/cm^3K^4\sim  \atop \rho _{mns}T_{ns}^{-4}=3.44\times 10^{-35}g/cm^3K^4}%
\end{equation}

The agreement now is not perfect, because there is an indeterminacy in $\rho
_{mns}$, but it is certainly much better than that between Eqs. (18) and
(19).

As shown below, However, we can make use of the $^{4}He$ cosmic abundance to
refine the experimental estimate of $n$ in the equation of state prior to
decoupling.\vskip1pc {\bf VI. }$^{4}He${\bf \ cosmic abundance and the time
at nucleosynthesis} \vskip1pc The neutron to proton ratio $(n/p)$, which is
determined by the $^{4}He$ abundance at nucleosynthesis and became frozen
thereafter \cite{Olive} is given by

\begin{equation}
(n/p)\simeq \frac 12Y_P/(1-\frac 12Y_p)\simeq 0.131,
\end{equation}

$(Y_p=0.232\pm 0.008),$ combined with the neutron lifetime

\begin{equation}
\tau _n=14.78\min =887\sec .,
\end{equation}

leads to

\begin{equation}
{(n/p)\simeq e^{-t_{ns}/\tau _n}\ i.e. \atop t_{ns}=t(T_{ns})=1803\sec .}%
\end{equation}

For $y<<1$ the cosmological equations reduce to

\begin{equation}
t=(R_{+}/\left| k\right| ^{1/2}c)\frac 23y^3,\ y=(\rho /\rho _{+})^{-1/6}
\end{equation}

This, with $\rho T^n\sim cont.$, entails $tT^{n/2}=const.$.

At $t=t_{af}=1.47\times 10^{13}\sec $ and $T=T_{af}=3808$, from which, with $%
n=3$, we get

\begin{equation}
t_{af}T_{af}^{3/2}=3.45\times 10^{18}\sec K^{3/2},
\end{equation}

which is way off

\begin{equation}
t_{ns}T_{ns}^{3/2}=1.16\times 10^{16}\sec K^{3/2},
\end{equation}

Letting n free in $tT^{n/2}\simeq const$ (equivalent to $\rho T^{-n}\simeq
const.$), on the other hand, we get

\begin{equation}
n=2\frac{\ln (t_{af}/t_{ns})}{\ln (T_{ns}/T_{af})}=3.99
\end{equation}

which gives a perfect agreement with $n=4$. Now we get

\begin{equation}
t_{af}T_{af}^{2}=2.1\times 10^{20}\simeq t_{ns}T_{ns}^{2}=2.1\times
10^{20}\sec K^{2}
\end{equation}

This is consistent with $\rho _{ns}=\rho _{af}(t_{af}/t_{ns})^{2}\simeq
0.125g/cm^{3}$, about one fourth of the value used above.

Thus fusion data and $^{4}He$ cosmic abundance data appear to reinforce each
other in support of an equation of state prior to decoupling

\begin{equation}
\rho _mT^{-4}\simeq \rho _{af}T_{af}^{-4}\simeq 8.47\times
10^{-36}g/cm^3K^4\ (t<t_{af})
\end{equation}

We may note that other light element abundances \cite{Coppi}, $(D,\ B,\ Be,\
^7Li)$, very minute in comparison with the $^4He$ abundance, and always open
to interpretation \cite{Burbidge}, are not considered here.

Let us go one step further and investigate physical conditions at the cosmic
threshold for baryons and beyond.\vskip1pc {\bf VII. Baryon threshold and
beyond} \vskip1pc Primordial elementary particles cannot be said to have
individual existence before the cosmic density $\rho _m$ was equal to the
particle density $\rho _z\simeq m_z/\frac{4\pi }3r_z^3$, where $m_z$ is the
mass of the particle and $r_z$ its radius. The threshold temperature for $%
\rho _m\simeq \rho _z$ is given by Stephan-Boltzmann's law as $%
T_z=m_zc^2/2.8k_B$.

For baryons (protons, neutrons) $m_b=1.67\times 10^{-24}g$ and $r_b\simeq
1.2\times 10^{-13}cm$ (from scattering experiments). With this data we get $%
\rho _b=2.0\times 10^{14}g/cm^3$ and $T_b=3.88\times 10^{12}K$ which gives

\begin{equation}
{\rho _bT_b^{-3}=3.42\times 10^{-24}g/cm^3K^3>> \atop \rho _{maf}T_{af}^{-3}=3.23\times 10^{-32}g/cm^3K^3}%
\end{equation}

On the other hand

\begin{equation}
{\rho _bT_b^{-4}=8.82\times 10^{-37}g/cm^3K^4\sim  \atop \rho _{maf}T_{af}^{-4}=8.82\times 10^{-36}g/cm^3K^4}%
\end{equation}

which is in relatively good order of magnitude agreement with an equation of
state $\rho _mT^{-4}\sim const.\ (\rho _m=\rho _r)$ prior to decoupling
(opaque universe). With this equation the baryon to photon ratio at $T=T_b$
is

\begin{equation}
\frac{n_b(T_b)}{n_r(T_b)}=\frac{\rho _b/m_b}{(\rho _rc^2)/2.8k_BT_b}=\frac{%
T_b}{m_bc^2/2.8k_B}=1
\end{equation}

being, consequently, $(n_b/n_r)_{T<T_b}$ less than one, and $%
(n_b/n_r)_{T>T_b}$ more than one. At $T_{eq}=3808K$ (atom
formation/decoupling), using Eq. (12) and (13), the baryon/photon ratio
becomes

\begin{equation}
\frac{n_b(T_{af})}{nr(T_{af})}=\frac{\rho _{meq}/m_b}{(\sigma
T_{eq}^4)/2.8k_BT_{eq}}=9.77\times 10^{-10}
\end{equation}

and, as noted in section III, it remains constant thereafter up to the
present epoch and beyond, as long as the universe remains transparent. In
other words, prior to decoupling, with the new equation of state (applicable
to an opaque universe), the baryon to photon ratio decreases as times goes
on, and the background temperature decreases accordingly.

In the meantime radiation pressure is continuously pushing matter radially,
until the time of last scattering (atom formation). At this time the
universe becames transparent, radiation pressure stops, protogalaxies begin
to form, and their constituent atoms, having already gained a tremendous
momentum from the background radiation in the plasma (opaque) phase of the
expansion, departed from each other at enormous speed. After decoupling
(atom formation) the universe has become transparent, the baryon to photon
ratio has become frozen, and consequently $%
(n_{b}/n_{r})_{T_{o}}=(n_{b}/n_{r})_{T_{af}}$.

For Planck's monopoles $m_p=(ch/G)^{1/2}=5.4\times 10^{-5}g$ and $%
r_p=(Gh/c^3)^{1/2}=4.0\times 10^{-33}cm$. With these numbers we obtain $\rho
_p=2.0\times 10^{92}g/cm^3$ and $T_p=3.5\times 10^{32}K$, which result in

\begin{equation}
\rho _pT_p^{-4}=1.33\times 10^{-38}\sim \rho _{maf}T_{af}^{-4}=8.82\times
10^{-36}g/cm^3,
\end{equation}

again in relatively good order of magnitude agreement with an equation of
state $\rho _mT^{-4}\sim const.\ (\rho _m=\rho _r)$. Cosmic time at this
temperature,by means of Eq. (29) is given by

\begin{equation}
{t_p=t_{af}(T_{af}/T_p)^2=1.74\times 10^{-45}\sim  \atop (Gh/c^5)^{1/2}=1.33\times 10^{-43}\sec .}%
\end{equation}

We can go a final step back in time. The earliest conceivable time
determined by Heisenberg's uncertainty principle for a universe of mass $%
M\sim \rho _{+}\frac{4\pi }3R_{+}^3\sim 5\times 10^{55}g$ is

\begin{equation}
t=t_L=h/Mc^2\simeq 1.46\times 10^{-103}\sec .
\end{equation}

This time can be properly called Lemaitre's time $(t_L)$, and a single
particle of this mass, a Lemaitre monopole (the ''primordial atom'' or
''primitive egg'' of George Lemaitre). The Compton radius of this monopole
would be $r_L=h/Mc=4.4\times 10^{-93}cm$. With these numbers we get $\rho
_L=4.3\times 10^{333}g/cm^3$ and $T_L=1.1\times 10^{92}K$, resulting in

\begin{equation}
\rho _LT_L^{-4}=2.9\times 10^{-35}\sim \rho _{maf}T_{af}^{-4}=8.82\times
10^{-36}g/cm^3,
\end{equation}

which is not in bad order of magnitude agreement with our equation of state
for the opaque universe prior to decoupling.\vskip1pc {\bf VIII. Concluding
remarks} \vskip1pc As we have seen, the gross features of cosmic dynamics
seem to be reasonably well described with a twofold equation of state

\begin{equation}
\rho _mT^{-4}=\sigma /c^2=8.47\times 10^{-36}g/cm^3K^4,\ at\ t<t_{af}
\end{equation}

and

\begin{equation}
\rho _mT^{-3}=(\sigma /c^2)T_{af}=3.23\times 10^{-32}g/cm^3K^3\ at\ t>t_{af}
\end{equation}

being $t_{af}=1.47\times 10^{13}\sec =0.466\times 10^6yrs$ and $T_{af}=3808K$%
, respectively, the time and temperature for atom formation (decoupling).
After decoupling, $\rho _mT^{-3}=const.$ is the standard, currently accepted
equation of state. Prior to decoupling $\rho T^{-4}\simeq const.$ implies a
drastic departure\cite{Gonzalo} from standard usage.

Eqs. (39) and (40) contain only universal constants and $T_{af}$, directly
related to $\hbar \omega _{af}=2.8k_BT_{af}$, the energy at which $H$ atoms
(the main constituents of the universe) begin to form, as given by Saha's
law, Eq. (14).

The proposed change may be unconvincing or even wrong. But one think it is:
simple.

One of us (J.A.G.) would like to thank S.L.Jaki for historical perspective
on the subject, and J.C.Mather and A.Hewish for correspondence and friendly
criticism of previous versions of this work. We thank H.Schulz for pointing
out to us more accurate physical data at the sun%
\'{}%
s center.

\end{document}